# LogStamping: A blockchain-based log auditing approach for large-scale systems

Md Shariful Islam and M. Sohel Rahman[1]
Department of CSE, BUET, ECE Building, West Palashi, Dhaka-1205

## Abstract

Log management is crucial for ensuring the security, integrity, and compliance of modern information systems. Traditional log management solutions face challenges in achieving tamper-proofing, scalability, and real-time processing in distributed environments. This paper presents a blockchain-based log management framework that addresses these limitations by leveraging blockchain's decentralized, immutable, and transparent features. The framework integrates a hybrid on-chain and off-chain storage model, combining blockchain's integrity guarantees with the scalability of distributed storage solutions like IPFS. Smart contracts automate log validation and access control, while cryptographic techniques ensure privacy and confidentiality. With a focus on real-time log processing, the framework is designed to handle the high-volume log generation typical in large-scale systems, such as data centers and network infrastructure. Performance evaluations demonstrate the framework's scalability, low latency, and ability to manage millions of log entries while maintaining strong security guarantees. Additionally, the paper discusses challenges like blockchain storage overhead and energy consumption, offering insights for enhancing future systems.

## 1. Introduction

Log management is crucial for IT operations, providing critical insights for monitoring, troubleshooting, and ensuring compliance. Various IT infrastructure components, including but not limited to servers, firewalls, routers, switches, and individual PCs, typically create log recordings (popularly referred to as 'logging'), particularly when they carry out crucial operations and transactions. Such logs are crucial for determining the trail of illicit actions carried out in such contexts [32, 33, 34]. Thus, these log files are often utilized to audit the computing environment, and they present us with crucial evidence for locating and resolving various incorrect or malicious behaviours that are interfering with the system under consideration and the infrastructure thereof. Even though log data can be useful in many ways, it can also be manipulated [37] to conceal harmful activities or impede the discovery of system vulnerabilities. Therefore, creating a secure and immutable system to store the vast amount of log data is essential to guarantee the integrity and safety of the computing environment.

[1] Corresponding Author: msrahman@cse.buet.ac.bd

Blockchain [35] is a shared, distributed, immutable ledger that facilitates the process of recording transactions and tracking assets, where an asset may refer to anything of value, both tangible and intangible. For its immutable and other desirable properties, blockchain has been utilized [38, 39, 40] in this context to create tamper-proof log record storage. Unfortunately, the continuously expanding huge log data cannot be handled efficiently by the current blockchain-based solutions [17, 19], which results in massive storage overhead [42, 43, 44, 45, 46] on the participating blockchain nodes, which in turn affects performance severely, sometimes compromising the original goal of integrity as well. While some works use separate off-chain storage to solve the storage scalability problem [15, 17], these works are unable to ensure the confidentiality of log data, thereby compromising a crucial issue [15, 23].

The adoption of blockchain for log auditing in large-scale systems is hindered by limitations in scalability, performance, and compliance. Traditional blockchain architectures suffer from low throughput and high storage overhead—storing $S$ GB of logs across $N$ nodes results in $S * N$ GB of total storage, making it impractical for terabyte-scale, high-frequency environments. Additionally, consensus mechanisms introduce latency, impacting the feasibility of real-time log recording. Current solutions inadequately handle the dynamic and incremental nature of log generation. Many assume that log files are secured at the source when generated, but it fails to address scenarios where logs are continuously appended, altered during transmission, or modified before ingestion. This oversight creates vulnerabilities in log integrity and auditability. Furthermore, waiting time during log generation—a critical factor in real systems—is often ignored, leading to unaddressed latency and potential data gaps. Some methods restructure logs for query efficiency at the cost of provenance, original format, and timestamp accuracy, which undermines forensic and compliance objectives. Energy-intensive mechanisms, such as PoW exacerbate inefficiency, and the immutability of blockchain leads to data redundancy and storage overhead, worsening scalability issues.

With this backdrop, this paper proposes a blockchain-based framework for log management that ensures tamper-proof logging, supports real-time processing, and maintains scalability in large-scale distributed environments. By leveraging smart contracts, cryptographic techniques, and a layered architectural design, the framework also prioritizes privacy and efficient recovery mechanisms.

# 2. Related Works

The advancement of log management systems, coupled with their integration into blockchain technology, has marked significant milestones and yielded valuable research contributions.

Various blockchain-based approaches have been developed to address similar challenges, with some solutions excelling in storing limited-size critical data [18, 21]. Other blockchain-based systems primarily focus on the storage and querying of logs directly from the blockchain [15, 41]. While each approach is well-suited to its specific use case, none effectively addresses the demands of large-scale systems with massive volumes of log data.

Tamper-resistant log files are essential in various domains and are mandated by numerous regulatory frameworks and standards, including HIPAA [47], and GDPR [48]. The integrity issue of these files is critical with varying degrees from one domain to another. For example, medical records must be reliable due to their potential life-or-death implications, financial data requires accuracy to maintain trust, and IT security logs are indispensable for detecting security incidents and conducting forensic investigations. A shared characteristic of these use cases is that log files are typically append-only, with new entries continuously added over time as individuals undergo more medical procedures, perform additional financial transactions, or generate further security events.

Beyond ensuring integrity, these logs must also ensure high availability to allow users to review and access records as and when needed. They serve crucial roles in fault analysis [1], anomaly detection [2, 3], forensic investigations [36], audits, and other critical processes [4, 10, 11, 12]. During a security breach, attackers often attempt to erase event logs on compromised systems to conceal their activities, underscoring the importance of secure and immutable log storage to preserve critical information and enhance system resilience.

Olaf and Esmiralda [4] proposed a centralized log server that can collect and store log records securely. However, this approach is vulnerable due to a single point of failure and lacks efficient query mechanisms. Indrajit et al. [5] introduced a cloud-based log storage system, but issues of trust and data consistency remained, as cloud servers are susceptible to unauthorized access and manipulation. A blockchain-based solution for immutable log storage was proposed in [6] that incorporated hierarchical ledgers to address scalability issues. While promising, the system, LogChain, lacks implementation details, and its API is underdeveloped for production-level deployments. Kumar et al. proposed a high-level design for secure log storage leveraging blockchain and cloud infrastructure [7]. However, the solution lacks details regarding its operational framework, performance evaluation, and query mechanisms.

Blockchain has also been explored in the domain of cloud forensics. Liang et al. [8] introduced ProvChain, a blockchain-based architecture for validating cloud data provenance, while Park et al. [9] proposed a data logging and integrity verification system for cloud environments. Both systems focused on cloud data integrity but failed to ensure log data integrity. Moreover, they did not provide a real-time performance analysis. Schneier and Kelsey pioneered cryptographic support for secure logs, emphasizing tamper detection in untrusted machines, laying the groundwork for tamper-proof logging [10]. However, this work lacked scalability, which is  essential for large-scale distributed systems.

Holt introduced Logcrypt, which enhanced log integrity through forward security and public verification, addressing critical gaps in audit log systems [11]. The major limitation was its dependency on centralized systems, which made it prone to single points of failure. Ahmad et al. presented BlockAudit, leveraging blockchain's immutability for secure and transparent audit logs, showcasing improved security and fault tolerance [12]. A drawback of this approach was its reliance on high storage overhead on-chain, which limited its scalability. Notably, IBM highlighted blockchain's storage challenges, advocating for efficient on-chain and off-chain strategies to handle growing data volumes [13].

Rakib et al. proposed [14] a MultiChain-based system for storing, querying, and auditing network logs. Their work achieves immutability, confidentiality, and scalability but focuses primarily on timestamp-based queries and does not emphasize real-time applicability to large-scale environments. Ali et al. introduced BCALS [15], a blockchain-based secure log management system tailored for cloud computing, ensuring audit log immutability and trust enhancement. The system's scalability was limited in handling diverse and high-frequency log sources. Furthermore, it transforms the logs before storing them into the blockchain, which creates a crucial concern with regards to the originality of the log. Shekhtman and Waisbard developed EngraveChain [16, 17], which leverages Hyperledger Fabric [27] to provide tamper-proof log storage with encryption for data privacy. However, it lacks efficient query mechanisms and comprehensive performance evaluations, particularly in large-scale systems. Rakib et al. further refined blockchain-enabled scalable network log systems, leveraging IPFS [49] for efficient data management and a robust query mechanism [19]. While it improves scalability as off-chain storage helps reduce on-chain data, the blockchain still maintains transaction metadata, which can lead to scalability concerns as the number of log transactions grows over time.

Collectively, these works underscore the potential of blockchain technology to address critical challenges in log management systems, including tamper-proofing, scalability, and privacy. However, challenges related to log confidentiality, real-time processing, and handling large log files remain as research gaps motivating further research and development.

# 3. Background

## 3.1 Blockchain

A blockchain is a decentralized and distributed ledger technology that securely records transactions across a network of computers. Transactions are grouped into blocks, each cryptographically linked to its predecessor, forming an immutable chain. This structure ensures transparency, security, and tamper-proof storage, making blockchain ideal for applications such as cryptocurrency, supply chain management, and smart contracts. Blockchain's core features include decentralization, immutability, transparency, and cryptographic security [59], enabling efficient, trustless operations across various industries.  Blockchain platforms provide the infrastructure for building, deploying, and managing decentralized systems and applications. They enable recording, validating, and securing data in an immutable, distributed ledger.

Several blockchain platforms are popular at the industry scale due to their unique capabilities and applications. For instance, Ethereum [26] is well-suited for private and consortium blockchains in enterprises, leveraging the Proof of Authority [52] consensus for fast block creation without mining. It supports smart contracts and decentralized applications (dApps) [55] within Ethereum's robust ecosystem. Hyperledger Fabric [27], another notable platform, is widely used in supply chain, finance, and healthcare industries. Its modular, permissioned architecture with private channels allows high customization for specific business workflows. Similarly, Corda [54] is designed for financial services and trade finance, featuring a peer-to-peer transaction model that ensures privacy and compliance with regulatory requirements.

Quorum [56], a blockchain platform forked from Ethereum, is tailored for banking and asset management, offering enhanced privacy and compatibility with Ethereum smart contracts. MultiChain [57], on the other hand, is designed for private networks and secure data sharing, providing fast deployment and built-in permissions management, making it ideal for controlled enterprise environments. Lastly, Ripple (XRP Ledger) [58] focuses on cross-border payments, delivering near-instant transactions and scalable performance for financial institutions. These platforms collectively address diverse enterprise needs, offering strong privacy, scalability, and customizability to support a wide range of business applications, from secure data sharing to financial services and decentralized asset management.

## 3.2 Smart contract

A smart contract is a self-executing program stored on a blockchain, with the terms and logic encoded directly into its code. The contract automatically executes when predefined conditions are met, ensuring tamper-proof, transparent, and trustless operations without intermediaries. Key features include automation, immutability, cryptographic security, and decentralized execution. Smart contracts are extensively used in financial transactions, supply chain management, and decentralized applications (dApps), transforming how agreements are enforced securely and efficiently.

## 3.3 Consensus algorithms

A consensus algorithm is a fundamental mechanism in blockchain networks that ensures all participants (nodes) agree on the validity of transactions and the current state of the ledger. It resolves trust issues in decentralized systems by providing a unified agreement among distributed nodes. **Table 1** compares three widely used consensus algorithms—PoW, PoA, and BFT—highlighting their trade-offs in terms of security, scalability, decentralization, and efficiency.

***Table 1*** *: Comparison of PoW, PoA, and BFT consensus mechanisms based on key features like security, scalability, energy efficiency, and use cases*

| Feature | Proof of Work (PoW) | Proof of Authority (PoA) | Byzantine Fault Tolerance (BFT) |
|---|---|---|---|
| **Security** | High | Medium | High |
| **Energy Efficiency** | Low | High | Medium |
| **Scalability** | Low | High | Medium |
| **Decentralization** | High | Medium | Medium |
| **Fault Tolerance** | Medium | Low | High |
| **Use Cases** | Bitcoin, Litecoin | VeChain, Rinkeby | Hyperledger, Tendermint |

### 3.4 InterPlanetary File System (IPFS)

The InterPlanetary File System (IPFS) [49] is a decentralized, peer-to-peer file storage and sharing protocol designed to create a more open and resilient web. Unlike traditional centralized systems, IPFS uses content-addressing to identify files by their unique cryptographic hash rather than their location. This ensures data integrity and allows files to be distributed across multiple nodes globally, enhancing reliability and resistance to censorship. IPFS is commonly used for storing and sharing large datasets, decentralized applications (dApps), and blockchain-related data, providing an efficient, secure, and scalable alternative to traditional file storage systems

### 3.5 Elasticsearch

Elasticsearch [60] is a distributed, open-source search and analytics engine built on Apache Lucene. It provides fast and scalable full-text search, data indexing, and real-time data exploration, making it ideal for applications like log analysis, business intelligence, and security monitoring. Its ability to handle large datasets efficiently makes it a popular choice for enterprise solutions.

## 4. Methods

In this section, we describe our research and experimental design in detail, discussing the rationale behind our design choices. We use the following technologies in our research:

a. Ethereum (Proof of Authority)

b. Solidity for smart contracts
c. IPFS for off-chain storage
d. Elasticsearch for search and analytics

## 4.1 Ethereum as our Blockchain Platform

Ethereum is one of the most widely used and versatile blockchain platforms, making it an excellent choice for developing secure, scalable, and decentralized applications. Its robust ecosystem offers extensive developer tools, active community support, and compatibility with smart contracts via the Ethereum Virtual Machine (EVM). These attributes make Ethereum particularly suited for enterprise-grade solutions and research applications.

## 4.2 Smart contracts with Ethereum (PoA)

The performance of a blockchain platform mostly depends on the consensus algorithm employed therein. The combination of Ethereum's versatile blockchain capabilities and PoA's high efficiency creates an optimized environment for scalable and secure applications. This configuration ensures rapid transaction processing, reduced operational costs, and robust smart contract execution, making it a preferred choice for enterprise and research-focused projects. We use Solidity with Ethereum (PoA) for its native EVM compatibility, enabling efficient, secure, and low-latency execution of smart contracts in a permissioned environment.

## 4.3 IPFS for Off-Chain Storage

Once a log file is verified using our tool and no longer changes, it is stored in the InterPlanetary File System (IPFS) to make it tamper-proof and persistently available. IPFS uses a unique hash to identify each file, ensuring its integrity. We then record that hash on the blockchain, creating a lightweight and verifiable audit trail. A similar approach was adopted by Rakib et al. [13], showcasing the use of IPFS for securely storing finalized logs in blockchain-based systems. However, their solution is confined to offline or pre-generated logs and does not address the challenges of real-time log generation, ingestion, or on-the-fly verification—key requirements for dynamic and continuously operating environments.

## 4.4 Elasticsearch for Search and Analytics

Querying logs directly from the blockchain is slow and not suitable for large-scale systems. To solve this, we use Elasticsearch for fast and efficient access to logs after they are verified and stored in IPFS. Finalized logs are indexed, making it easy to search, filter, and analyze them quickly. This setup keeps integrity checks handled by the blockchain and IPFS, while Elasticsearch ensures fast performance for tasks like audits, anomaly detection, and compliance.

## 4.5 Data Model

In this study, we focus on plain-text logs, where each log entry follows a standardized format to ensure uniformity and compatibility. A typical log entry includes:

- **Timestamp**: Precise date and time with nanosecond granularity to maintain accuracy.
- **Log Level**: Indicates the severity or priority of the log (e.g., INFO, DEBUG, ERROR).
- **Machine/Service Name**: Specifies the source of the log for identification.
- **Log Details**: Provides a description or message for the logged event.

This standardized structure enables efficient parsing, storage, and analysis of log data, crucial for large-scale systems.

## 4.6 Data Collection

To comprehensively test the system's performance, both offline and real-time, we utilize three distinct sources of log data as follows.

1) **Online Archives**: We have collected datasets from LogPai [13], which contain diverse log samples from large-scale systems and data centers. These datasets allow us to evaluate the system's offline behavior with substantial volumes of data.
2) **Synthetic Log Generators**: We have used tools like Fake-Apache-Log-Generator [14] to generate human-readable, randomized logs. These enable us to simulate diverse scenarios and test the system's real-time data handling capabilities.
3) **Custom Log Generator**: A tailored log generation tool has been developed to create logs with specific patterns, formats, or parameters. This process ensures flexibility for testing system behaviors under customized conditions.

By combining these datasets, we aim to rigorously evaluate the system's robustness, scalability, and real-time processing capabilities across a range of scenarios and data volumes.

# 5. Main Approach of LogStamping

We developed our system with three major components, they are:

1) Ingestion Tool,
2) Blockchain Platform,
3) and Integrity Verification Tool.

In what follows, we briefly describe these components.

## 5.1 The Ingestion Tool

The Python-based log ingestion tool is designed to provide a scalable and secure solution for managing log data in large-scale systems. By integrating blockchain technology, the tool ensures the immutability, traceability, and auditability of log entries. It continuously monitors log files for new entries, generates cryptographic hashes (using **SHA256 [50]**) for individual or a group of *n* log lines, where *n* is any predefined number of log lines, and records these hashes on the blockchain. This approach ensures that log data remains tamper-proof and can be reliably audited for compliance and forensic purposes if and when required. The tool is particularly suited for high-volume environments, such as data centers and distributed systems, where traditional log management systems often struggle to maintain security and scalability. The following components make up the modular architecture of the log ingestion tool:

- **Log Collector**: Gathers logs from various sources.
- **Parser and Formatter**: Standardizes log formats.
- **Blockchain Interface**: Interacts with the blockchain to store logs immutably.
- **Error Handling Module**: Manages exceptions and logging failures.

**Algorithm 1**: **MonitorAndIngestLogs**
**Input**: logFile, groupSize, timeout
**Output**: Hashes stored in blockchain for log file integrity

```
1. Initialize logGroup ← ∅
2. Initialize startTime ← CURRENT_TIME()

3. while True do
4.    Wait for new log entry in logFile
5.    if NEW_ENTRY_EXISTS(logFile) then
6.       Append log entry to logGroup
7.    end if

8.    if |logGroup| ≥ groupSize or (CURRENT_TIME() − startTime) ≥ timeout then
9.       hashValue ← GenerateSHA256Hash(logGroup)
10.      WriteToBlockchain(hashValue)
11.      logGroup ← ∅
12.      startTime ← CURRENT_TIME()
13.   end if
14. end while
```

**Algorithm 2: GenerateSHA256Hash**
**Input**: logGroup
**Output**: Hash value of log group

1. Concatenate all log entries in logGroup into a single string
2. return SHA256_*HASH*_OF_STRING(string)

**Algorithm 3: WriteToBlockchain**
**Input**: hashValue
**Output**: hashValue stored in blockchain

1. Connect to blockchain
2. Store hashValue in blockchain
3. return SUCCESS

The tool works by executing the following steps. **Figure 1** illustrates the described log ingestion process.

1. **Monitoring Log Entries**: The log ingestion tool continuously monitors the target log file for new entries, processing them line by line.

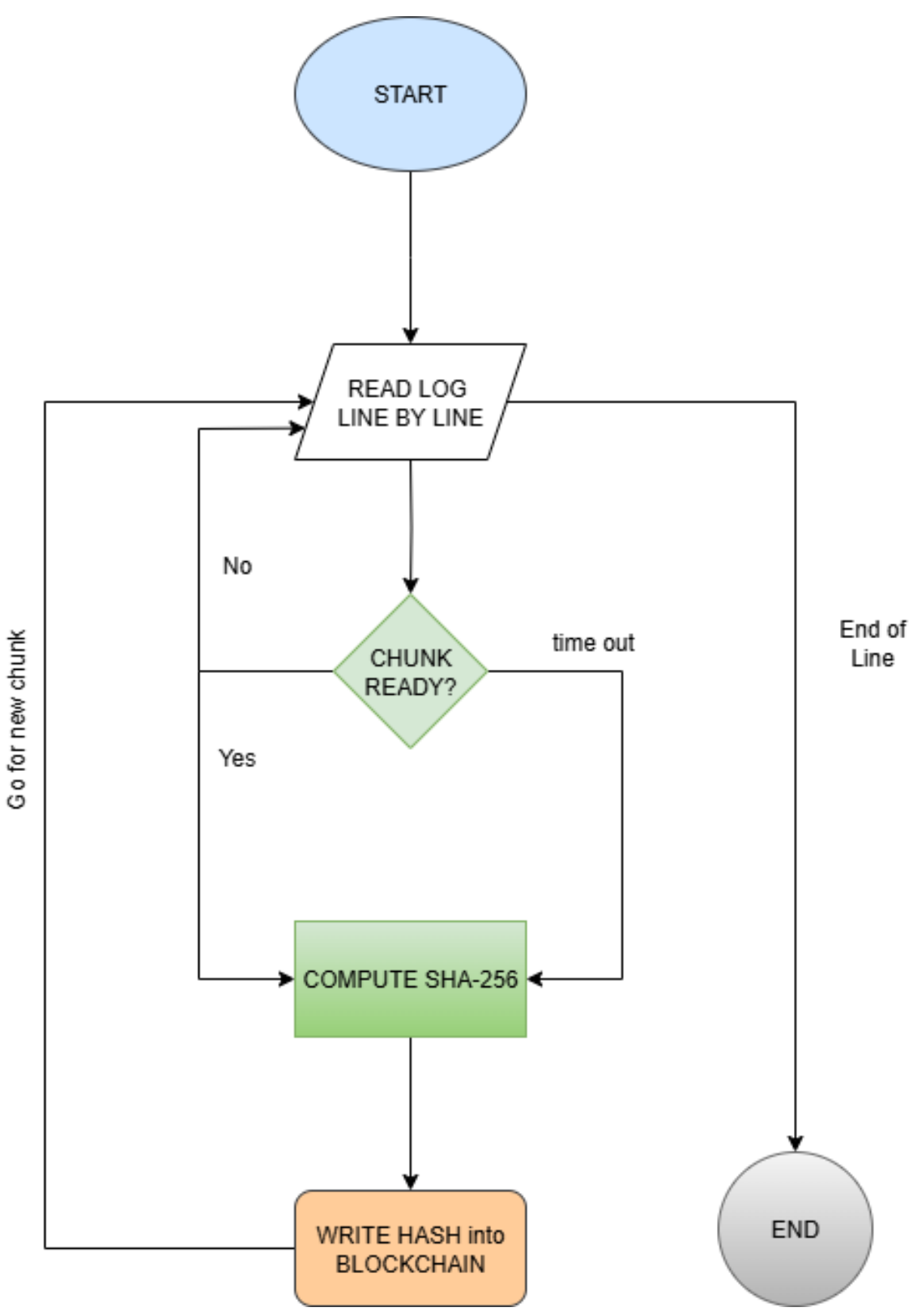

*Figure 1: Log ingestion workflow for secure blockchain logging using SHA-256 and group-based entry hashing.*

2. **Grouping Log Lines**: Instead of sending each log line individually, the tool groups multiple log lines to form a chunk based on pre-configured parameters (e.g., number of lines (*n)*, time intervals *(t)*). Here, Algorithm 1 explains the complete process.
   - **Timeout Handling**:
     i. If the chunk is incomplete (e.g., insufficient new log entries), the tool waits for a pre-configured timeout *(t)* period.
     ii. After the timeout, the hash of the partial group is computed and written to the blockchain to secure any unrecorded entries.
   - **Dynamic Group Capacity**:
     i. The size of the groups is variable, dynamically adjusting based on the frequency of log entries in the target log file.
3. **Hash Generation**: Once a group is formed, the tool computes SHA256 hash for the grouped log lines, creating a unique digital fingerprint (Algorithm 2).
4. **Writing to Blockchain**: The computed hash is immediately written into the blockchain, ensuring the immutability and integrity of the grouped logs (Algorithm 3).

## 5.2 The Blockchain Platform

We deployed Ethereum [26] as the private blockchain platform for its ease of setup, maintenance, and scalability. To improve efficiency as per Figure 1, we adopted the Proof of Authority (PoA) consensus algorithm [52], which enables rapid block creation without the need for mining. Custom smart contracts were developed in Solidity [51], supported by additional tools, such as Blockchain Explorer [53], for enhanced functionality. The smart contract's functionalities are in Algorithm 4.

**Algorithm 4: LogStorage Smart Contract**
**Purpose**: To store and retrieve log hashes securely on the blockchain.
**Input**: logHash (string) - The cryptographic hash of a log group.
**Output**: Immutable storage and retrieval of log hashes.

**State Variables:**
1. logHashes: A mapping (integer → string) to store log hashes indexed by their count.
2. logCount: An unsigned integer representing the total number of stored log hashes.

**Functions:**

1. Function: storeLogHash
   Input: logHash (string) - The hash to be stored.
   Output: Updates logHashes and increments logCount.

   Procedure:
   1. logHashes[logCount] ← logHash
   2. logCount ← logCount + 1

```
2. Function: getLogHash
   Input: index (unsigned integer) - The index of the log hash to retrieve.
   Output: The log hash stored at the specified index.

   Procedure:
   1. return logHashes[index]
```

The Integrity Verification Tool

Verification tools ensure the integrity and authenticity of log files by detecting tampering or modifications. Their process logs by grouping entries based on predefined parameters, computing cryptographic hashes using SHA256, and verifying these hashes against blockchain records. Using timestamps from log entries, they align verification with real-world events. Optionally, verified logs can be archived in IPFS for immutability or indexed in ElasticSearch for efficient querying. These tools are vital for audits, forensic investigations, and maintaining trust in system logs.

```
Algorithm 5: VerifyLogIntegrityWithTimestamps
Input: logFile, groupParameters (maxLines, maxWaitTime), blockchain, ipfs (optional),
elasticSearch (optional)
Output: Verification status of log integrity

1. Initialize logGroup ← Ø
2. Initialize allHashesValid ← TRUE
3. Initialize groupStartTime ← NULL
4. Initialize groupEndTime ← NULL

5. Open logFile for reading
6. while not EOF(logFile) do
7.    Read logLine from logFile
8.    Append logLine to logGroup
9.    Extract timestamp from logLine

10.   if groupStartTime = NULL then
11.      groupStartTime ← timestamp
12.   end if
13.   groupEndTime ← timestamp

14.   if |logGroup| ≥ groupParameters.maxLines or (groupEndTime − groupStartTime) ≥
groupParameters.maxWaitTime then
15.      hashValue ← ComputeHash(logGroup)
16.      isValid ← QueryBlockchain(hashValue, blockchain)

17.      if isValid = FALSE then
18.         allHashesValid ← FALSE
19.         Print "Tampered group detected:"
```

```
20.       end if

21.       logGroup ← Ø
22.       groupStartTime ← NULL
23.       groupEndTime ← NULL
24.    end if
25. end while

26. if logGroup ≠ Ø then
27.    hashValue ← ComputeHash(logGroup)
28.    isValid ← QueryBlockchain(hashValue, blockchain)

29.    if isValid = FALSE then
30.       allHashesValid ← FALSE
31.       Print "Tampered group detected:", logGroup
32.    end if
33. end if

34. if allHashesValid = TRUE then
35.    Print "Log file is intact"
36.    if ipfs ≠ NULL then
37.       ArchiveToIPFS(logFile)
38.    end if
39.    if elasticSearch ≠ NULL then
40.       StoreInElasticSearch(logFile)
41.    end if
42. else
43.    Print "Log file has been modified"
44. end if
```

The tool works based on Algorithm 5 by executing the following steps. **Figure 2** illustrates the log verification process.

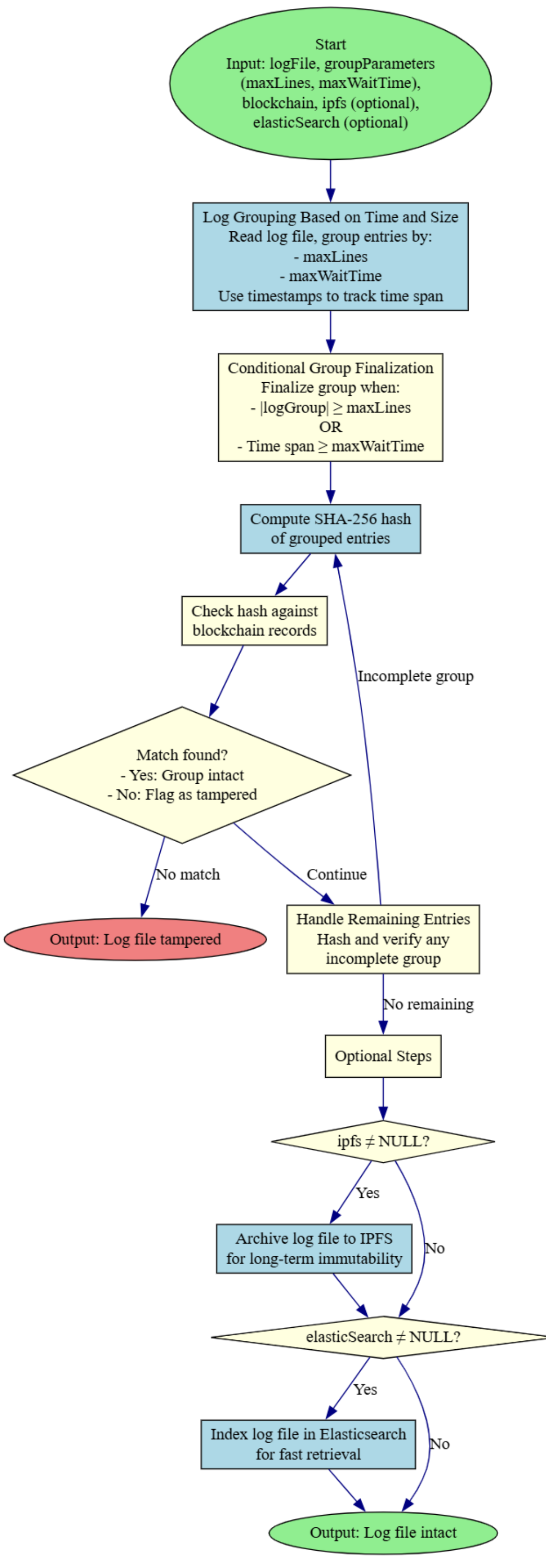

***Figure 2***: *Verification workflow that ensures log integrity by grouping original log entries and validating them against trusted records stored on the blockchain to identify any unauthorized modifications.*

1. **Log Grouping Based on Time and Size:** The algorithm reads an existing log file and groups entries based on two conditions: a maximum number of lines (**maxLines**) or a maximum time window (**maxWaitTime**). Each entry is added to the current group, and timestamps are used to determine the time span (Line 14).
2. **Conditional Group Finalization:** When either condition is met, the current group is finalized for integrity verification (Lines 14-20). This approach ensures consistent and adaptive log grouping without requiring real-time monitoring.
3. **Hash Computation and Blockchain Verification:** For every completed group -
    - A SHA-256 hash is computed from the grouped entries (Line 15, 27).
    - The hash is checked against blockchain records (Lines 34-44).
        - **Match found:** Group is confirmed intact.
        - **No match:** Group is flagged as tampered.
4. **Handling Remaining Entries:** After processing all log lines, any incomplete group is also hashed and verified to ensure no entries are skipped (Lines 26-33).
5. **Optional Steps:**
    - **IPFS Archival:** Verified log files can be archived in IPFS to ensure long-term immutability.
    - **Elasticsearch Indexing:** Logs can be indexed in Elasticsearch for fast retrieval and advanced search capabilities.

## 5.3 Handling log files in large scale systems

Efficient log management in large-scale systems relies on robust strategies for categorization, traceability, and organization. A widely adopted approach is the use of structured naming patterns for log files, incorporating base names dynamically configured with details, such as timestamps, system identifiers, or unique indices. This method ensures better organization, simplified retrieval, and improved log file management.

Common log generation methods include time-based, size-based, index-based, event-based, chunk-based, hybrid (time + size), and distributed approaches. These methods segment logs by criteria, such as time, size, events, indices, or sources, creating unique file names that often include timestamp fields for each entry.

Our system employs a hybrid approach (i.e., a combination of chunk-based and time-based approaches) to process real-time logs into the blockchain, leveraging standardized log generation patterns. The chunk-based approach groups logs by a fixed number of entries, ideal for high-volume systems to optimize performance and reduce processing overhead. On the other hand, the time-based approach groups logs by fixed time intervals, ensuring timely processing and enhanced security in systems with irregular log generation. Hybrid approach, as the name indicates, combines both chunk size and time interval conditions, finalizing log groups when either threshold is met for balanced efficiency and flexibility. This methodology ensures scalable, flexible, and reliable management practices optimized for real-time operations in large-scale systems.

To elaborate, in this approach, logs are grouped based on two criteria: a predefined maximum chunk size (number of entries) and a predefined time interval (e.g., seconds or minutes). A log group is finalized and processed as soon as either of these conditions are met, ensuring timeliness and scalability. This dual criterion prevents excessive waiting for logs to fill a chunk while avoiding overloading the system during high activity periods. Once a group is complete, a cryptographic hash is computed and stored on the blockchain, ensuring the integrity and traceability of the logs. The hybrid approach also allows dynamic adjustment of chunk size and time intervals, enabling the system to adapt to changes in log generation rates and workloads. By combining the strengths of both methods, the hybrid approach reduces tampering risks, optimizes resource usage, and ensures timely log processing, making it ideal for large-scale, high-frequency systems.

## 5.4 Log Archiving Using Off-Chain

In this approach, logs are periodically archived off-chain after the ingestion process for a specific log file is complete and no additional entries are expected. Once a log file is marked as complete, the verification tool continuously monitors it to ensure there are no alterations or tampering. If the verification tool confirms the file's integrity, the entire log file is encrypted using a symmetric key to enhance security and then archived in IPFS (InterPlanetary File System). This ensures that the archived file is both immutable and secure.

The archiving process is designed to adapt to the log generation strategy. For instance, if the strategy is time-based, the system will trigger the archiving process at regular time intervals, ensuring an encrypted and immutable copy of the original log file is maintained. This approach is beneficial for log recovery and for pinpointing specific lines where modifications might have occurred. Alternatively, strategies such as size-based or index-based log generation are also supported. Regardless of the strategy, the archiving process remains consistent, ensuring securely stored logs that are readily available for verification and recovery.

By encrypting the original file before storing it in IPFS, this off-chain archiving approach maintains log integrity and enhances confidentiality. This method provides a robust mechanism for safeguarding logs, supporting flexible log generation strategies, and ensuring immutable, tamper-proof, and secure records.

## 5.5 Leveraging Elasticsearch for Efficient Log Search and Audit

Searching data directly on a blockchain is not optimal, particularly for large-scale log files. Large data storage in a distributed manner requires a significant amount of storage capacity. Blockchain's inherent design prioritizes immutability and security but lacks the performance capabilities required for efficient data retrieval, especially for unstructured data. To address this limitation, industry-standard tools like Elasticsearch [60] are a better fit for full-text search and analysis. Elasticsearch is known for its high performance and scalability, making it ideal for handling large datasets and conducting fast, precise searches.

In our system, we utilized Elasticsearch for storing and querying logs after their integrity was verified by the verification tool. Logs are stored in chunks, with each chunk containing the following items.

1. **Calculated Hash**: Ensuring that data integrity is maintained and verifiable.
2. **Raw Log Data**: Providing unstructured log content for search and analysis.
3. **Chunk Metadata**: Including the hash of the chunk and its associated log data for additional traceability.

For logs originating from IPFS (InterPlanetary File System), Elasticsearch acts as a complementary storage solution. IPFS ensures the integrity, availability, and immutability of the log files, while Elasticsearch facilitates efficient full-text search and audit processes. This dual approach enhances both data security and retrieval performance.

The primary purpose of Elasticsearch in this system is to support forensic and auditing operations. By enabling fast and accurate searches across large datasets, Elasticsearch simplifies the task of finding specific log entries, even within unstructured data. This approach combines the security of blockchain and IPFS with the performance capabilities of Elasticsearch, creating a robust solution for log management in large-scale systems.

# 6. Experimental setup

## 6.1 Hardware Requirements

For the implementation of our system, we configured three identical nodes with the following hardware specifications:

- **Processor**: 4 vCPUs
- **Memory (RAM)**: 8 GB
- **Storage**: 200 GB
- **Operating System: Ubuntu** 22.04 LTS

These nodes are uniformly configured to ensure consistent performance across the blockchain network. This identical setup minimizes variations in processing and storage.

## 6.2 Datasets

For the evaluation of our system, we utilized two distinct datasets to assess performance under varying log volumes:

1) **Small Dataset**:
    a. **Size**: 10,000 log lines
    b. **Purpose**: Used to evaluate the correctness of our proposed model.
    c. **Datasource**: LogPai [13]
2) **Large Dataset**:
    a. **Size**: 14 million log lines (~1.3 GB)
    b. **Purpose**: Used to test the system's scalability and robustness in handling large-scale log data efficiently.
    c. **DataSource**: Fake-Apache-Log-Generator [14] and our custom log generator

These datasets provided comprehensive insights into the system's performance across both small-scale and large-scale use cases, ensuring its suitability for diverse operational requirements.

## 6.3 Tools

We use the following tools in our experiments:

- **Python:** Core language for implementing log processing, verification logic, and system integration.
- **web3.py:** Python library for interacting with Ethereum-compatible blockchains, handling smart contract interactions and transactions.
- **web3.js:** JavaScript library for blockchain communication from web or Node.js applications.
- **Geth:** Go Ethereum client used to run a full Ethereum node and interface with the blockchain.
- **IPFS CLI / API:** Tools for decentralized storage and retrieval of verified log files.
- **Elasticsearch:** Engine for indexing and querying verified logs efficiently.

## 6.4 Analysis

**Security**: The proposed model incorporates robust security measures to ensure data integrity and protection. By leveraging the Proof of Authority (PoA) consensus algorithm, it ensures that only trusted, pre-authorized nodes are responsible for block creation, minimizing the risk of unauthorized activity. Furthermore, the ingestion tool operates within a secure, private network, restricting access exclusively to verified entities. These combined mechanisms create a highly secure and reliable framework for log management.

**Scalability**: The proposed system enhances the scalability of blockchain networks by significantly reducing the number of network calls through the implementation of chunk-based processing. This approach minimizes the frequency of transactions, making the system more efficient and suitable for integration with other services or applications. Additionally, the reduced storage requirements decrease the input/output (I/O) overhead on blockchain servers, thereby improving processing efficiency and overall system performance.

**Privacy**: We have implemented robust privacy measures in our proposed model to ensure the highest level of data security. Notably, no raw log data is stored directly on the blockchain, and it is impossible to reconstruct raw data from the information stored in the blockchain. Furthermore, the data stored in the InterPlanetary File System (IPFS) is encrypted before being transmitted, adding an additional layer of security. This comprehensive approach ensures that privacy is maintained at an optimal level, adhering to best practices for secure and private data management.

## 6.5 Evaluation

We have evaluated the proposed system under various scenarios to analyze its behavior in terms of storage usage and performance. These scenarios were designed to assess the system's capabilities for both small and large log datasets. The experiments were conducted with different configurations, focusing on the time required for processing and storage consumption. **Table 2** reports the datasets and chunk parameters used in different experiments.

*Table 2: Experimental setup for evaluating log ingestion and storage efficiency, showing different dataset sizes, chunk configurations, log formats (raw and hashed), number of nodes, and the performance metrics (time and storage) used for analysis.*

| Dataset (log entries) | Chunk Size (N) | Ingestion Date Type | Nodes | Metrics |
|---|---|---|---|---|
| 10,000 | 1, 5, 10, 20 | Raw | 3 | Time & Storage |
| | | Hashed | | |
| 14 million | 1, 5, 10, 20 | Hashed | 3 | Time & Storage |

The storage required to store raw data on the blockchain is consistently high, regardless of the chunk size. Therefore, we excluded raw data from the analysis for large datasets, focusing instead on the results for different chunk sizes with hashed data.

# 7. Result Analysis & Discussions

## 7.1 Result Analysis

### 7.1.1 Ingestion Anlysis

We initially conducted experiments on a small dataset (10k log entries) to analyze storage usage for raw data and hashed data for each log line.

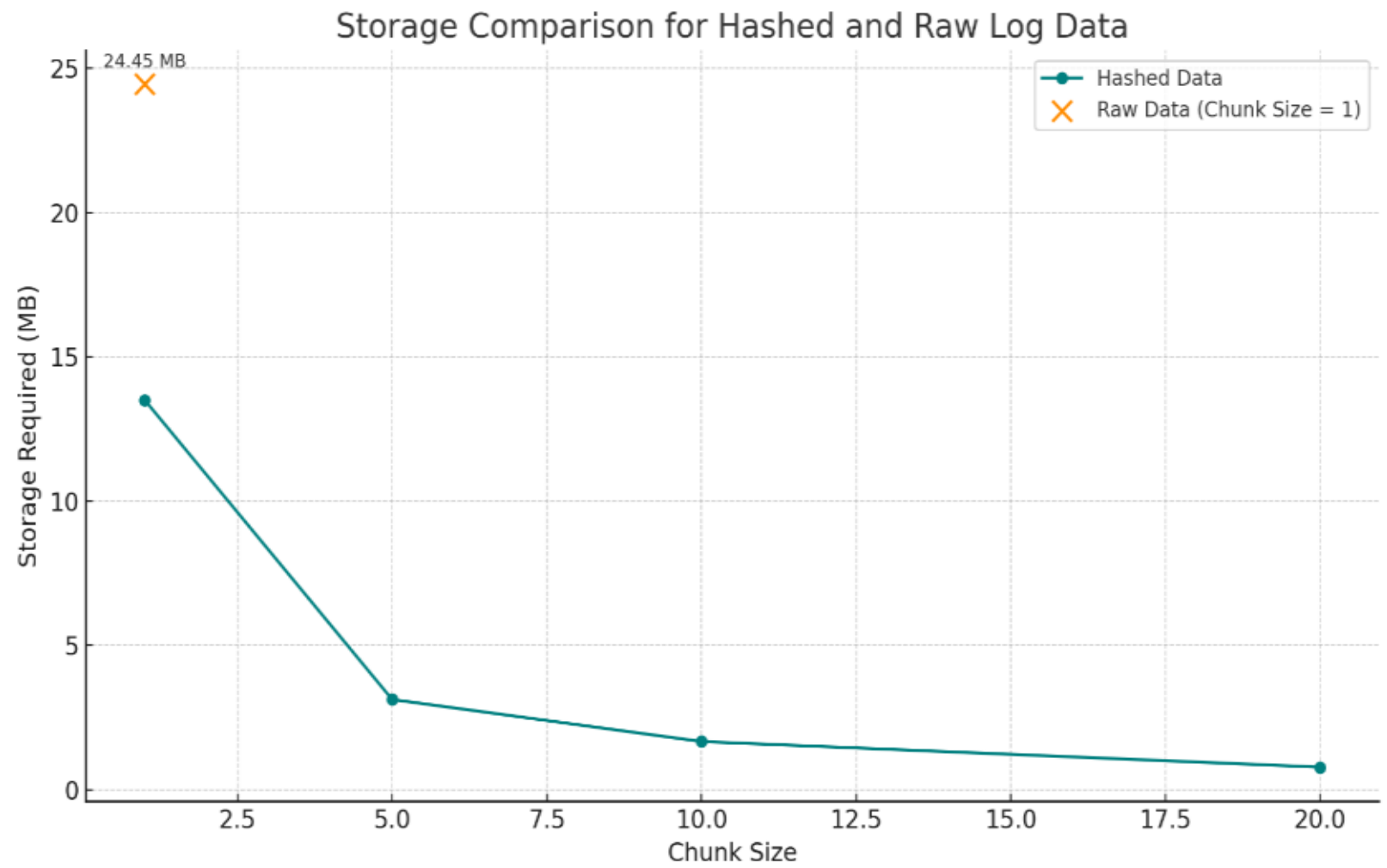

*Figure 3: Effect of Chunk Size on Storage Efficiency - Comparing Raw vs. Hashed Storage Overhead for 10,000 Log Entries*

From the results depicted in **Figure 3**, it is evident that for a chunk size of $N=1$, the storage required for hashed data is reduced by half compared to raw data. This reduction corresponds to a **50%** storage gain, highlighting the efficiency of hashed data storage in minimizing storage overhead while preserving data integrity and auditability.

The storage required for hashed data significantly decreases as the chunk size increases. For $N=5$, the storage usage is approximately **five times less** compared to $N=1$. This substantial reduction is due to the minimized metadata overhead and fewer chunks being generated, demonstrating the efficiency of larger chunk sizes in optimizing storage.

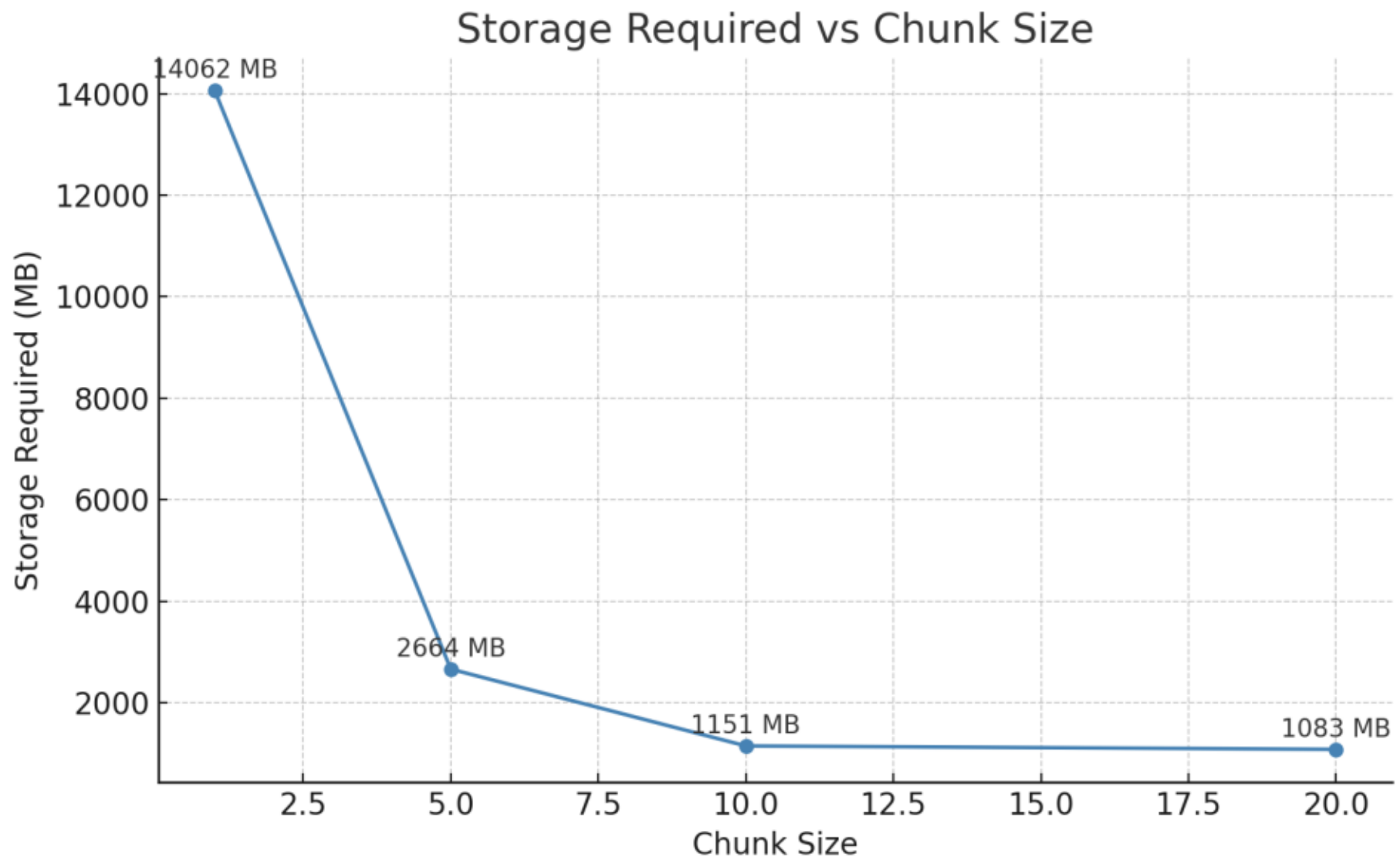

*Figure 4: Impact of Chunk Size on Storage Usage for 14 Million Log Entries*

From **Figure 4**, the storage required is *14 GB* where *N=1* and the data is hashed. It also shows that increasing the chunk size to *N=5, 10, 20* for hashed data significantly reduces storage requirements. For each increase in chunk size, the storage requirement becomes approximately half of the same for the preceding chunk size. This reduction occurs due to fewer chunks being created, which minimizes metadata overhead and optimizes storage usage. Larger chunk sizes are thus highly effective in reducing storage demands while maintaining data integrity.

Similarly, Figure 5 presents a quantitative analysis of processing time for 14 million hashed log entries under varying chunk sizes. When each log entry is processed individually (chunk size of 1), the total processing time reaches approximately **172 hours**, indicating a substantial computational overhead. However, when logs are grouped into batches of 5 entries, the processing time drops dramatically to about **16 hours**, reflecting a nearly **90% reduction**. Increasing the chunk size further results in additional gains—processing time decreases to roughly **7.5 hours** with a chunk size of 10 and stabilizes around **7.3 hours** at chunk size 20.

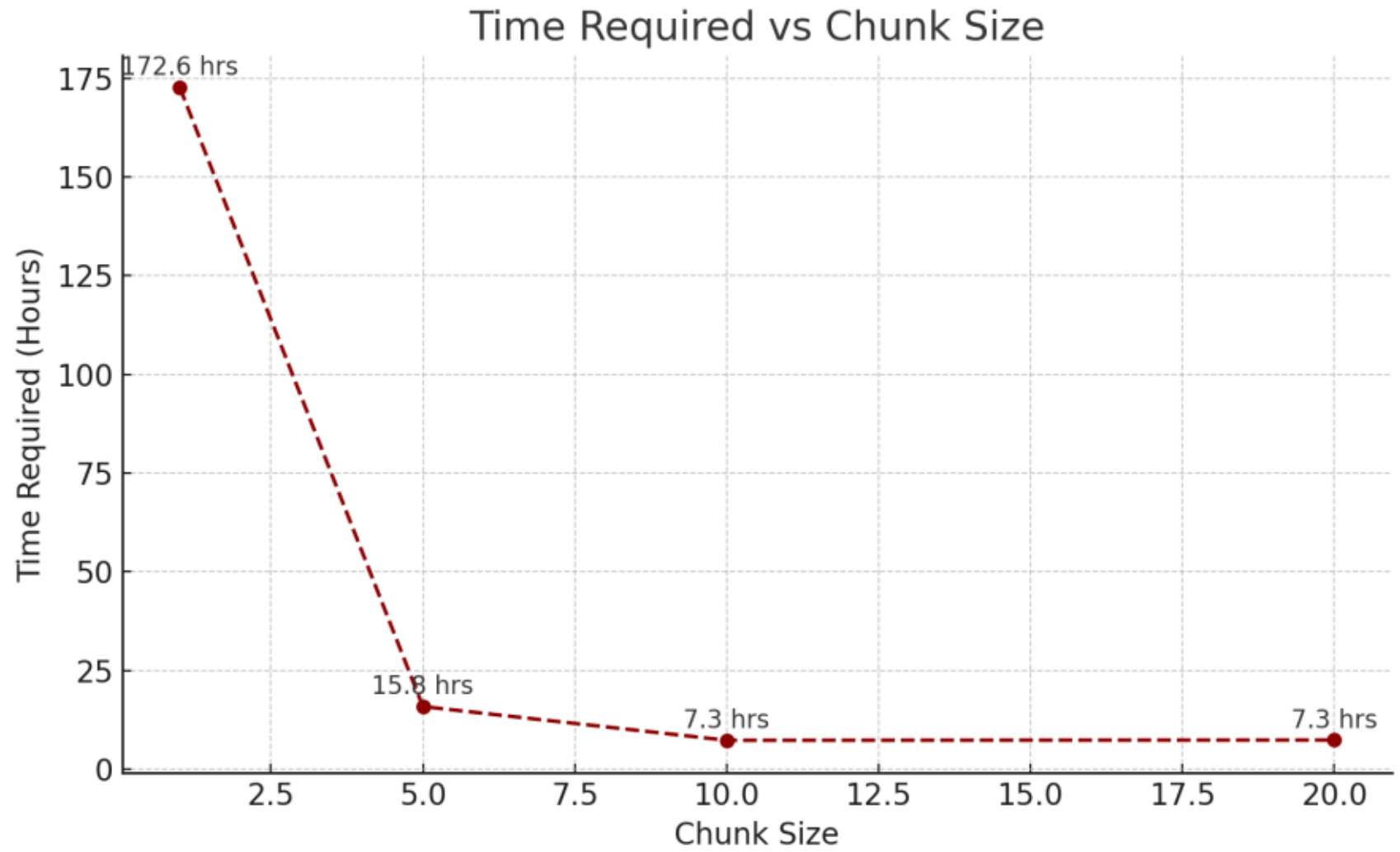

*Figure 5: Processing Time for Hashed Log Data Decreases Significantly as Chunk Size Increases, Highlighting the Efficiency Gains in Batch-Based Log Handling*

This trend clearly demonstrates that batching log entries into larger groups significantly improves processing efficiency. Fewer hash computations and reduced storage interactions contribute to the time savings. These findings emphasize the value of selecting an appropriate chunking strategy to ensure scalability and high performance in large-scale log processing systems. **Figure 4 and Figure 5** collectively offer details about the scalability and performance of the system, emphasizing its ability to handle large-scale datasets efficiently in terms of both storage and time.

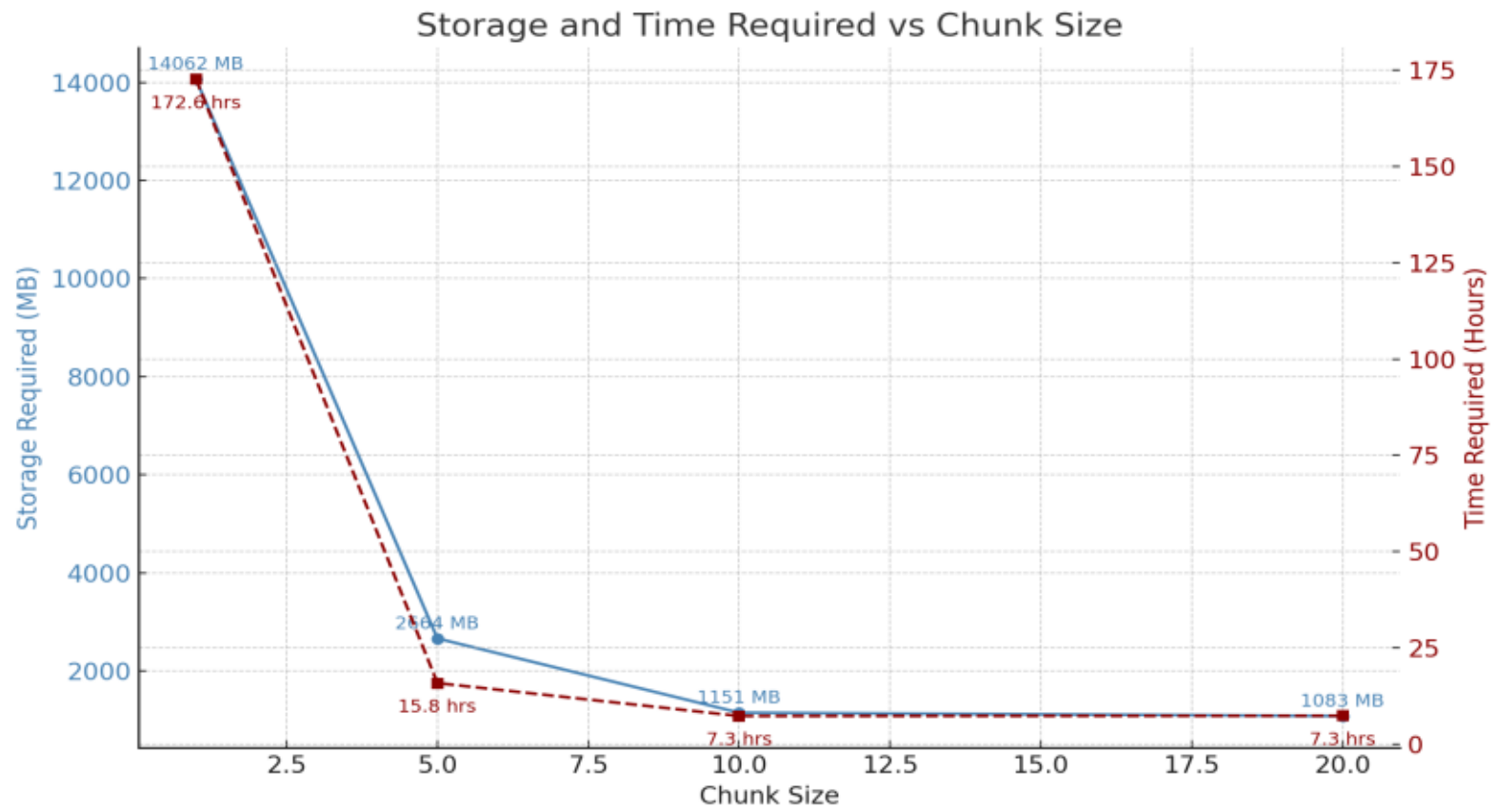

*Figure 6: Impact of Chunk Size on Storage and Processing; larger chunks significantly reduce both storage overhead and processing duration during log data handling. (14 Million Entries)*

Finally, **Figure 6** presents a combined analysis of storage usage and processing time for 14 million log entries using various chunk sizes. This figure demonstrates how increasing the chunk size significantly reduces both storage overhead and processing time during log data processing. When the chunk size is set to 1, the system requires approximately **14,062 MB** of additional storage and takes around **172.6 hours** to process, highlighting the inefficiency of handling logs individually. As the chunk size increases to 5, both metrics improve drastically, with storage dropping to **2,664 MB** and processing time reducing to **15.8 hours**. This trend continues with chunk size 10, where storage falls to **1,151 MB** and time to **7.3 hours**. At chunk size 20, the gains begin to plateau, with storage at **1,083 MB** and processing time at **7.3 hours**, indicating diminishing returns beyond this point. Overall, Figure 6 demonstrates that batching logs into larger chunks significantly enhances system efficiency, with the most notable improvements occurring between chunk sizes 1 and 10.

These results clearly demonstrate that increasing the chunk size drastically reduces both storage requirements and processing time. The greatest gains occur between chunk sizes 1 and 5, emphasizing the inefficiency of handling logs individually. The trend highlights the effectiveness of chunking for optimizing both system performance and resource utilization in large-scale log management.

### 7.1.2 Verification Analysis

The verification time analysis presented in **Figure 7** highlights the significant impact of chunk size (N) on the efficiency of log verification. When the chunk size is set to 1, the system takes approximately **1,020 minutes** (or 17 hours) to verify the logs, indicating a high computational burden when each entry is processed individually.

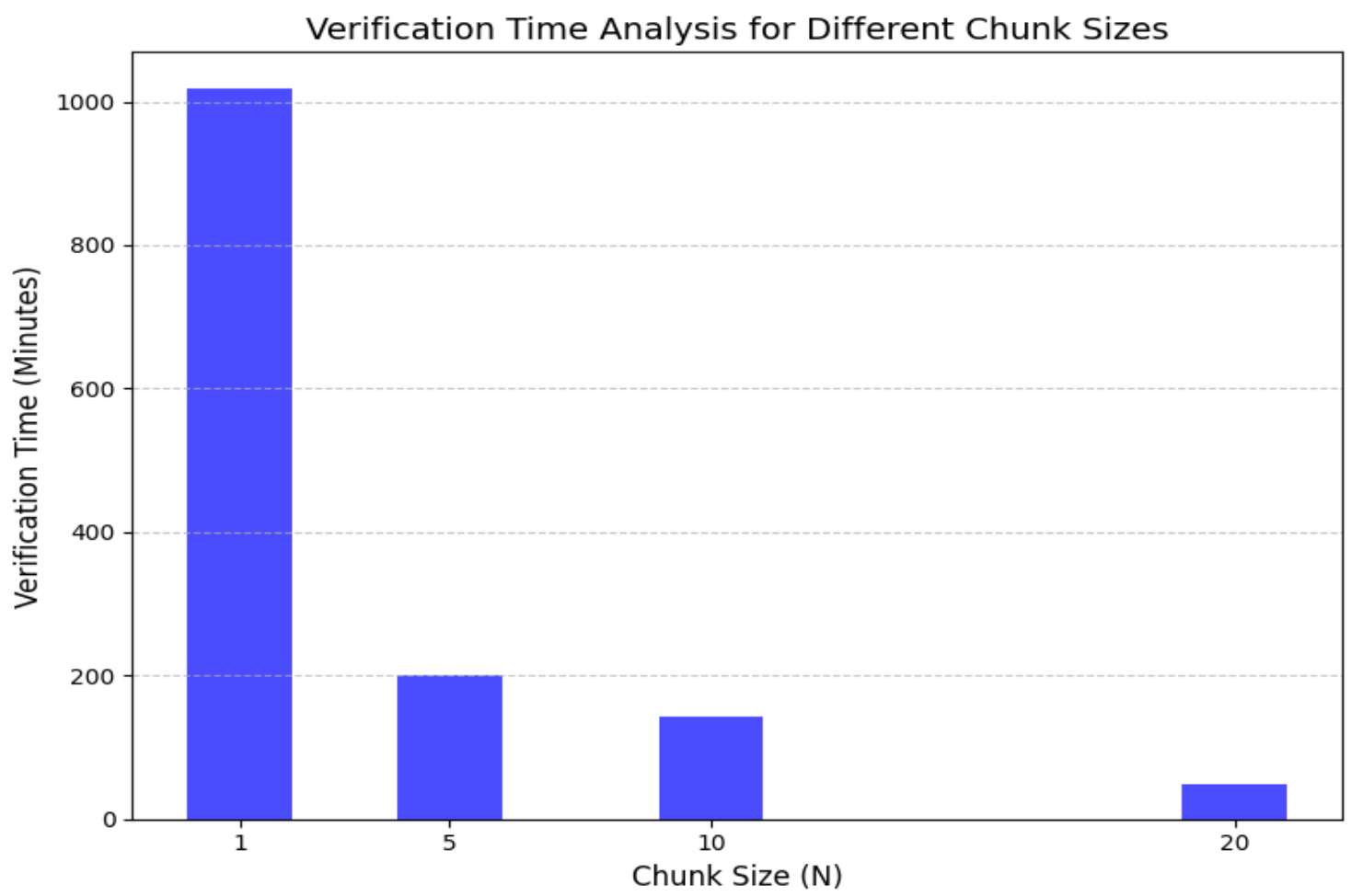

*Figure 7: Verification time decreases sharply with larger chunk sizes, dropping from 1,020 minutes at size 1 to just 50 minutes at size 20—demonstrating up to 95% efficiency gain in log verification.*

However, as the chunk size increases to 5, the verification time drops sharply to **200 minutes**, representing nearly an **80% reduction**. At chunk size 10, the time further decreases to around **145 minutes**, and at chunk size 20, it reaches just **50 minutes**. Overall, moving from a chunk size of 1 to 20 results in a **95% reduction** in verification time. This trend highlights the efficiency benefits of processing logs in larger chunks, especially in large-scale or real-time verification systems.

These results highlight the importance of optimizing chunk sizes for efficient log verification. Larger chunk sizes reduce processing overhead and enhance the scalability of the system, particularly in high-volume log datasets. By balancing chunk size with processing capabilities, this approach ensures a more efficient and scalable solution for log verification in large-scale systems.

## 7.2 Comparative Analysis Summary

This section compares **LogStamping** against four notable blockchain-based log or provenance systems: **LogChain [6]**, **BCALS [15]**, **EngraveChain [16]**, **Rakib et al. [19]**, and **ProvChain[8]**. The analysis evaluates key system aspects including storage, verification speed, real-time capability, query performance, and tamper detection. **Table 3** shows a comparative analysis summary among those systems.

***Table 3****: Comparison of Blockchain-Based Log Management Systems*

| **Feature** | **LogChain** | **BCALS** | **EngraveChain** | **Rakib et al.** | **ProvChain** | **LogStamping(Proposed)** |
| --- | --- | --- | --- | --- | --- | --- |
| **Storage Efficiency** | Unknown: No backend specified | Moderate: On-chain; no compression | High: Encrypted chunks, no IPFS | High: IPFS + hash mapping | Low: Fully on-chain; high cost | ✅ Very High: IPFS + chunking (92% reduction) |
| **Verification Time** | Unreported: No results | Moderate: On-chain transforms slow audits | Moderate: Limited benchmarking | Moderate: No numbers reported | Moderate: No latency data | ✅ Very High: 95% faster audit (1,020 → 50 mins) |
| **Real-Time Support** | ❌ Not implemented; underdeveloped API | ❌ Limited | ❌ Limited | ❌ Pre-computed logs only | ❌ Not emphasized; post-event focus | ✅ High |

| **Query Efficiency** | Unknown: Query system not defined | Moderate: Uses Elasticsearch, not optimized | Moderate: Slow chain traversal | High: Timestamped Merkle tree | Low: No retrieval optimization | ✅ Very High: Indexed search via IPFS + Elasticsearch |
| --- | --- | --- | --- | --- | --- | --- |
| **Tamper Detection** | Theoretical: Not validated | High: Blockchain integrity | High: AES/RSA, no detection rate given | High: IPFS-backed hash proofs | High: Immutable, but no tested attack detection | ✅ Very High: 100% detection of fake log injections |

## 7.3 Discussion on Storage Gain Perspective

The analysis of storage requirements for different chunk sizes reveals significant storage optimization as the chunk size increases. For *N=1*, the storage demand is the highest due to the large number of small chunks, each requiring its own metadata and hash computations. This additional overhead contributes to inflated storage usage. In contrast, larger chunk sizes (*N=5, 10, 20*) dramatically reduce storage needs by consolidating more log entries into fewer chunks, thereby minimizing metadata overhead.

For instance, Figure 4 shows that when *N=5*, the storage requirement is approximately **five times less** than that for *N=1,* demonstrating a substantial storage gain. Similarly, as the chunk size increases to *N=10 and N=20,* the storage demands continue to decrease, albeit with diminishing returns. This trend underscores the efficiency of larger chunk sizes in reducing overall storage requirements while maintaining data integrity.

From a storage optimization perspective, the results indicate that larger chunk sizes reduce the total storage footprint and make the system more scalable for large datasets. However, there is a trade-off between storage efficiency and potential delays in log ingestion, as larger chunks require more time to fill. Balancing chunk size based on system requirements and log generation rates is essential for achieving optimal storage usage and system performance. This approach highlights the importance of chunk size configuration as a critical parameter for scalable and efficient blockchain-based log management.

## 7.4 Discussion on Performance Gain Perspective

The performance of blockchain systems in processing large-scale log files is influenced by several critical factors, including the selection of the blockchain platform, the choice of an efficient consensus algorithm, and the underlying hardware capabilities, such as CPU performance and host machine specifications. After a thorough evaluation, we identified and implemented the most suitable platform combined with a high-performance consensus algorithm to optimize processing efficiency. Our findings highlight the time required to process large log files, demonstrating that our approach leverages these optimizations to achieve enhanced performance and scalability.

As previously discussed, Figure 5 highlights substantial performance gains achieved by increasing chunk size. Recall that processing time drops from **172.6 hours at N=1** to **15.8 hours at N=5** (a ~91% improvement), and further to **7.3 hours at N=10** (a ~54% gain over N=5). These results confirm that the majority of performance gains occur between N=1 and N=10. Moreover, this approach enables near real-time processing when chunk sizes are dynamically adjusted based on application-specific log generation rates.

Overall, our experiments highlight that while increasing chunk size dramatically boosts performance at first, especially from *N=1 to N=10*, the marginal benefits taper off beyond that. Selecting a moderate chunk size, such as N=10, offers a balanced trade-off between performance and resource usage in large-scale log processing. Importantly, this figure also illustrates that when log generation occurs in real time, this chunk-based strategy allows the system to process logs almost in real time. By dynamically adjusting chunk sizes based on the log generation rate of different applications, the system can maintain efficiency and responsiveness without introducing significant processing delays.

## 7.5 Discussion on Temper detection

The proposed tamper detection mechanism leverages a timestamp-aware log grouping strategy to ensure data integrity. Using configurable thresholds (**maxLines**, **maxWaitTime**), log entries are grouped and hashed incrementally, with each hash verified against the blockchain. This enables localized detection of modifications without reprocessing the entire log file. The algorithm supports real-time log integrity checks, efficiently identifying altered segments based on their temporal boundaries. If tampering is detected, the affected group is flagged, while intact logs are optionally archived to IPFS and indexed in Elasticsearch. Experimental results show a 100% detection accuracy for any synthetic tampering, confirming the system's effectiveness for scalable, fine-grained integrity verification in dynamic environments.

# 8. Conclusion and Future Works

The proposed **LogStamping** framework addresses key limitations in existing blockchain-based log management systems by introducing an efficient, scalable, and tamper-resilient solution. Through dynamic chunking, off-chain storage via IPFS, and real-time indexing with Elasticsearch, the system enables high-throughput log ingestion and efficient auditability. Empirical evaluations confirm significant gains in storage efficiency, performance, and tamper detection accuracy.

Furthermore, the system guarantees integrity through cryptographic hashing and blockchain anchoring, ensuring trustworthy and verifiable log records. The name **LogStamping** reflects its foundational design—mirroring **timestamping practices in PKI systems**, where data integrity, authenticity, and temporal validity are enforced via secure digital signatures. Similarly, LogStamping secures logs by embedding them into timestamped hash groups, offering strong guarantees of chronological accuracy, traceability, and audit readiness. This work lays a robust foundation for future advancements in decentralized and verifiable system log management.

However, a key limitation lies in the system's **inability to recover logs during ingestion**. Since only hashes are stored on-chain and raw logs are not retained until explicitly archived, any loss or deletion before ingestion completes will result in irreversible data loss. Although this enhances privacy and reduces storage burden, it limits real-time recoverability. Once logs are ingested and verified, however, they become tamper-proof and immutable, providing reliable records for audit and compliance.

Future work will focus on extending the system's **log recoverability**, **fault-tolerant ingestion pipelines**, and **fine-grained access control** for multi-tenant environments. Integrating anomaly detection, dynamic chunk reconfiguration, and decentralized identity for audit attribution will further enhance LogStamping's applicability in critical infrastructure and cloud-native systems.

# Declarations

## Declaration of Competing Interest

The authors declare that they have no known competing financial interests or personal relationships that could have appeared to influence the work reported in this paper.

## Funding

This research received no specific grant from any funding agency in the public, commercial, or not-for-profit sectors.

## Data Availability

The data that support the findings of this study are available from the corresponding author upon reasonable request.

## Declaration of Generative AI and AI-assisted Technologies in the Writing Process

The authors acknowledge the use of OpenAI's ChatGPT to enhance the clarity, sentence structure, and readability of the manuscript. All scientific content, ideas, and original contributions were developed by the authors and thoroughly reviewed to ensure accuracy and integrity.

# References

1. Liang, Y.; Zhang, Y.; Sivasubramaniam, A.; Jette, M.; Sahoo, R. Bluegene/l failure analysis and prediction models. In Proceedings of the International Conference on Dependable Systems and Networks, Philadelphia, PA, USA, 25–28 June 2006; p. 425.
2. Frei, A.; Rennhard, M. Histogram Matrix: Log File Visualization for Anomaly Detection. In Proceedings of the 2008 Third International Conference on Availability, Reliability and Security, Barcelona, Spain, 4–7 March 2008; pp. 610–617.
3. Goldstein, M.; Raz, D.; Segall, I. Experience Report: Log-Based Behavioral Differencing. In Proceedings of the 2017 IEEE 28th International Symposium on Software Reliability Engineering (ISSRE), Toulouse, France, 23–26 October 2017; pp. 282–293.
4. O. Soderstrom and E. Moradian, "Secure audit log management," ¨ Procedia Comput. Sci., vol. 22, pp. 1249–1258, 2013
5. I. Ray, K. Belyaev, M. Strizhov, D. Mulamba, and M. Rajaram, "Secure logging as a service-delegating log management to the cloud," IEEE Syst. Journal, vol. 7, no. 2, pp. 323–334, 2013.
6. W. Pourmajidi and A. Miranskyy, "Logchain: Blockchain-assisted log storage," in Proc. 11th IEEE Int. Conf. Cloud Comput. (CLOUD), 2018, pp. 978–982.
7. M. Kumar, A. K. Singh, and T. V. S. Kumar, "Secure log storage using blockchain and cloud infrastructure," in Proc. 9th IEEE Int. Conf. Comput., Commun. and Netw. Technol. (ICCCNT), 2018, pp. 1–4
8. X. Liang, S. Shetty, D. Tosh, C. Kamhoua, K. Kwiat, and L. Njilla, "Provchain: A blockchain-based data provenance architecture in cloud environment with enhanced privacy and availability," in Proc. 17th IEEE/ACM Int. Symp. Cluster, Cloud and Grid Comput. (CCGrid'17), 2017, pp. 468–477
9. J. H. Park, J. Y. Park, and E. N. Huh, "Block chain based data logging and integrity management system for cloud **forensics**," Comput. Sci. & Inf. Technol., vol. 149, 2017
10. Schneier, B., & Kelsey, J. (1999). Cryptographic Support for Secure Logs on Untrusted Machines.
11. Holt, J. E. (2006). Logcrypt: Forward Security and Public Verification for Secure Audit Logs. AISW-NetSec 2006.
12. Ahmad, A., Saad, M., Bassiouni, M., & Mohaisen, A. (2018). Towards Blockchain-Driven, Secure and Transparent Audit Logs. MobiQuitous, 3286978-3286985.
13. Rakib, M. H., Hossain, S., Jahan, M., & Kabir, U. (2020). Towards Blockchain-Driven Network Log Management System. IEEE iSCI
14. IBM (2018). Storage Needs for Blockchain Technology: Point of View. IBM Corporation.
15. Ali, A., Khan, A., Ahmed, M., & Jeon, G. (2021). BCALS: Blockchain-Based Secure Log Management System for Cloud Computing. Transactions on Emerging Telecommunications Technologies, e4272.

16. L. M. Shekhtman and E. Waisbard, "Securing log files through blockchain technology," in Proceedings of the 11th ACM International Systems and Storage Conference, Article No. 3, 2018. doi: 10.1145/3211890.3211893.
17. Shekhtman, L., & Waisbard, E. (2021). EngraveChain: A Blockchain-Based Tamper-Proof Distributed Log System. Future Internet, 13(6), 143.
18. P. V. Kakarlapudi and Q. H. Mahmoud, "Design and Development of a Blockchain-Based System for Private Data Management," Electronics, vol. 10, no. 24, p. 3131, 2021. doi: 10.3390/electronics10243131.
19. Rakib, M. H., Hossain, S., Jahan, M., & Kabir, U. (2022). A Blockchain-Enabled Scalable Network Log Management System. Journal of Computer Science, 18(6), 496-508.
20. Y. Zhao, X. Liu, and L. Wang, "A Complete Log Files Security Solution Using Anomaly Detection and Blockchain Technology," Proceedings of the IEEE International Conference on Big Data (Big Data 2023), pp. X-X, 2023. DOI: 10.1109/BigData.2023.10100200.
21. Kanhere, S. S., & Conti, M. (2024). Blockchain for Health Data Management. In Blockchains: A Handbook on Fundamentals, Platforms and Applications (pp. 321-346). Springer. https://doi.org/10.1007/978-3-031-32146-7_18
22. Khan, S., Alam, M., & Khan, S. U. (2023). Blockchain-Based Secure Logging Mechanism for Cloud Forensics. Computers & Security, 125, 102976. https://doi.org/10.1016/j.cose.2023.102976
23. Singh, A., Zhou, Y., Mehrotra, S., Sadoghi, M., Sharma, S., & Nawab, F. (2023). WedgeBlock: An off-chain secure logging platform for blockchain applications. In Proceedings of the 26th International Conference on Extending Database Technology (EDBT) (pp. 684–696). https://doi.org/10.48786/edbt.2023.43
24. Zhang, P., & Wang, J. (2019). Blockchain Based Data Integrity Verification for Large-Scale IoT Data. IEEE Access, 7, 164401-164411. https://doi.org/10.1109/ACCESS.2019.2952847
25. Nakamoto, S. (2008). Bitcoin: A peer-to-peer electronic cash system. Retrieved from https://bitcoin.org/bitcoin.pdf
26. Buterin, V. (2013). A next-generation smart contract and decentralized application platform. Ethereum Whitepaper. Available at https://ethereum.org/en/whitepaper/
27. Hyperledger Fabric Documentation. (2024). Hyperledger Fabric documentation. Retrieved from https://hyperledger-fabric.readthedocs.io/
28. A Large Collection of System Log Datasets for AI-Driven Log Analytics, https://github.com/logpai/loghub,[Last Access on 12 Jan 2024] Generate a Boatload of Fake Apache Log Files Very Quickly,
29. https://github.com/kiritbasu/Fake-Apache-Log-Generator, [Last Access on 12 Jan 2024].
30. Yoshida, H., & Biryukov, A. (2005). Analysis of a SHA-256 variant. In Advances in Cryptology – EUROCRYPT 2005 (pp. 245–260). Springer. https://doi.org/10.1007/11693383_17

31. Satoshi, Nakamoto . Bitcoin: A peer-to-peer electronic cash system, https://bitcoin.org/bitcoin.pdf, [Last Access on 12 Jan 2024].
32. Chuvakin, A., Schmidt, K., & Phillips, C. (2013). *Logging and Log Management: The Authoritative Guide to Understanding the Concepts Surrounding Logging and Log Management.* Syngress.
33. Kent, K., & Souppaya, M. (2006). *Guide to Computer Security Log Management (NIST Special Publication 800-92).* National Institute of Standards and Technology (NIST).
34. Barisani, A., & Oldani, D. (2007). *Offensive Computing: Understanding Windows, Linux, and UNIX Security.* Syngress.
35. Yaga, D., Mell, P., Roby, N., & Scarfone, K. (2018). Blockchain technology overview. National Institute of Standards and Technology (NIST). [https://doi.org/10.6028/NIST.IR.8202](https://doi.org/
36. J. Dykstra and A. T. Sherman, "Understanding issues in cloud forensics: two hypothetical case studies," in Proceedings of the Conference on Digital Forensics, Security and Law, 2011, p. 45.
37. Aptive, "Log Injection Attack: Understanding and Mitigating the Risks." [Online]. Available: https://www.aptive.co.uk/blog/log-injection-attack/. [Accessed: Jan. 12, 2024].
38. Ma, D., & Tsudik, G. (2009). A new approach to secure logging. ACM Transactions on Storage, 5(1), 2:1–2:21. https://doi.org/10.1145/1502777.1502779
39. Rajebhosale, S. S., & Nikam, M. C. (2019). Development of secured log management system over blockchain technology. International Journal of Cyber Research and Education, 1(1), 38–42. IGI Global. https://doi.org/10.4018/IJCRE.2019010104
40. Ahmad, A., Saad, M., Bassiouni, M., & Mohaisen, A. (2018). Towards blockchain-driven, secure and transparent audit logs. In Proceedings of the 15th EAI International Conference on Mobile and Ubiquitous Systems: Computing, Networking and Services (pp. 443–448). ACM. https://doi.org/10.1145/3286978.3286985
41. Xu, G., Yun, F., Xu, S., Yu, Y., Chen, X.-B., & Dong, M. (2023). *A blockchain-based log storage model with efficient query.* **Soft Computing, 27**, 13779–13787. https://doi.org/10.1007/s00500-023-08975-3
42. Khan, D., Jung, L. T., & Hashmani, M. A. (2020). Scalability in blockchain: Challenges and solutions. In Blockchain Technology: Applications and Challenges (pp. 315–332). Elsevier. https://doi.org/10.1016/B978-0-12-819816-2.00015-0
43. Shafin, K. M., & Reno, S. (2024). Breaking the blockchain trilemma: A comprehensive consensus mechanism for ensuring security, scalability, and decentralization. IET Software, 2024(1). https://doi.org/10.1049/2024/6874055
44. Abbas, Z., & Myeong, S. (2024). A comprehensive study of blockchain technology and its role in promoting sustainability and circularity across large-scale industry. Sustainability, 16(10), 4232. https://doi.org/10.3390/su16104232

45. Sanka, A. I., & Cheung, R. C. C. (2021). A systematic review of blockchain scalability: Issues, solutions, analysis and future research. Journal of Network and Computer Applications, 195, 103232. https://doi.org/10.1016/j.jnca.2021.103232
46. Alghamdi, T. A., Khalid, R., & Javaid, N. (2024). A survey of blockchain based systems: Scalability issues and solutions, applications and future challenges. IEEE Access, 12, 79626–79651. https://doi.org/10.1109/ACCESS.2024.3408868
47. U.S. Department of Health & Human Services. (1996). Health Insurance Portability and Accountability Act (HIPAA). Retrieved from https://www.hhs.gov/hipaa/index.html, [Last Accessed on 24 May 2024].
48. European Union. (2016). General Data Protection Regulation (GDPR). Retrieved from https://eur-lex.europa.eu/eli/reg/2016/679/oj, [Last Accessed on 24 May 2024].
49. Benet, J. (2014). *IPFS - Content Addressed, Versioned, P2P File System.* arXiv:1407.3561. DOI: 10.48550/arXiv.1407.3561.
50. National Institute of Standards and Technology (NIST), "Secure Hash Standard (SHS)," Federal Information Processing Standards Publication 180-4 (FIPS PUB 180-4), 2015. [Online]. Available: https://doi.org/10.6028/NIST.FIPS.180-4.
51. Ethereum Foundation, *Solidity: High-Level Language for Smart Contracts*, 2024. [Online]. Available: https://soliditylang.org.
52. V. Buterin, "Proof of Authority Chains," Ethereum Foundation, 2017. [Online]. Available: https://github.com/ethereum/EIPs/issues/225.
53. "Blockchain explorer," *Wikipedia, The Free Encyclopedia*, Apr. 2024. [Online]. Available: https://en.wikipedia.org/wiki/Blockchain_explorer.
54. Brown, R. G., Carlyle, J., Grigg, I., & Hearn, M. (2016). Corda: An introductory whitepaper. R3 Consortium. Retrieved May 24, 2024, from https://docs.r3.com/en/pdf/corda-introductory-whitepaper.pdf
55. Ethereum Foundation. (n.d.). *Introduction to dapps.* Retrieved May 24, 2024, from https://ethereum.org/en/developers/docs/dapps/
56. J.P. Morgan. (n.d.). *Quorum blockchain platform.* Retrieved May 24, 2024, from https://consensys.net/quorum/
57. MultiChain**.** (n.d.). *MultiChain: Open platform for building blockchains.* Retrieved May 24, 2024, from https://www.multichain.com/
58. Ripple Labs Inc. (n.d.). *Ripple: Real-time gross settlement system, currency exchange and remittance network.* Retrieved May 24, 2024, from https://ripple.com/
59. Zheng, Z., Xie, S., Dai, H., Chen, X., & Wang, H. (2017, June). *An overview of blockchain technology: Architecture, consensus, and future trends.* In 2017 IEEE International Congress on Big Data (BigData Congress) (pp. 557–564). IEEE. https://doi.org/10.1109/BigDataCongress.2017.85
60. Elastic. (n.d.). *Elasticsearch: Distributed, RESTful search and analytics engine.* Retrieved May 24, 2024, from https://www.elastic.co/elasticsearch/